\documentclass[journal,12pt,onecolumn]{IEEEtran}
\usepackage{float}
\usepackage{amssymb}
\usepackage[T1]{fontenc}
\usepackage{amsthm}
\usepackage{units}
\usepackage{etoolbox}
\usepackage[latin1]{inputenc}
\usepackage{amsmath}
\usepackage{amsfonts}
\usepackage{amssymb}
\usepackage{graphicx}
\usepackage{stmaryrd}
\usepackage{cite}
\usepackage{bm}
\usepackage{algorithmic}
\usepackage{array}
\usepackage{amstext}
\usepackage{latexsym}
\usepackage{color}
\usepackage{cite,url}
\usepackage{setspace}
\usepackage{hyperref}
\newtheoremstyle{note}
{3pt}
{3pt}
{}
{}
{\bfseries}
{:}
{.5em}
{}
\theoremstyle{note}
\newtheorem{lem}{Lemma}
\newtheorem{thm}{Theorem}

\makeatletter
\patchcmd{\@makecaption}
{\scshape}
{}
{}
{}
\makeatletter
\patchcmd{\@makecaption}
{\\}
{.\ }
{}
{}
\makeatother

\usepackage{pdflscape}
\allowdisplaybreaks
\usepackage{multirow}

\DeclareMathSizes{10}{9}{8}{7}

\begin{document}

\title{Multi-Antenna Channel Interpolation via Tucker Decomposed Extreme Learning Machine}

\author{Han~Zhang,~\IEEEmembership{Student Member,~IEEE}, Bo Ai,~\IEEEmembership{Senior Member,~IEEE}, Wenjun Xu,~\IEEEmembership{Senior Member,~IEEE}, Li~Xu,~\IEEEmembership{Member,~IEEE,} and Shuguang~Cui,~\IEEEmembership{Fellow,~IEEE}
	\thanks{Manuscript received February 24, 2019; revised April 05, 2019; accepted April 23, 2019; date of current version April 25, 2019. The work was supported in part by NSF with grants CNS-1824553, DMS-1622433, AST-1547436, and ECCS-1659025.}
	\thanks{Copyright (c) 2015 IEEE. Personal use of this material is permitted. However, permission to use this material for any other purposes must be obtained from the IEEE by sending a request to pubs-permissions@ieee.org. }%
	\thanks{H. Zhang and~S.~Cui are with the Department of Electrical and Computer Engineering, University of California, Davis, CA 95616, USA (e-mails: \{hanzh, sgcui\}@ucdavis.edu), Bo Ai is with the State Key Laboratory of Rail Traffic Control and Safety, Beijing Jiaotong University, Beijing 100044, China (e-mail: aibo@ieee.org), Wenjun Xu is with the Key Lab of Universal Wireless Communications, Ministry of Education, Beijing University of Posts and Telecommunications(e-mail: wjxu@bupt.edu.cn), and Li Xu is with the Institute of Computing Technology, Chinese Academy of Science, Beijing 100190, China(e-mail: lixu@ict.ac.cn).}}

\markboth{IEEE Transactions on Vehicular Technology,~Vol.~XX, No.~XX, XXX~2015}
{}
\maketitle

\begin{abstract}
Channel interpolation is an essential technique for providing high-accuracy estimation of the channel state information (CSI) for wireless systems design where the frequency-space structural correlations of multi-antenna channel are typically hidden in matrix or tensor forms. In this letter, a modified extreme learning machine (ELM) that can process tensorial data, or ELM model with tensorial inputs (TELM), is proposed to handle the channel interpolation task. The TELM inherits many good properties from ELMs. Based on the TELM, the Tucker decomposed extreme learning machine (TDELM) is proposed for further improving the performance. Furthermore, we establish a theoretical argument to measure the interpolation capability of the proposed learning machines. Experimental results verify that our proposed learning machines can achieve comparable mean squared error (MSE) performance against the traditional ELMs but with 15\% shorter running time, and outperform the other methods for a 20\% margin measured in MSE for channel interpolation.
\end{abstract}


\IEEEpeerreviewmaketitle

\section{Introduction}

 High-quality channel estimation is crucial for many wireless applications, which is resource demanding in both time and frequency, where channel interpolation \cite{interpol1} and prediction \cite{Cpredictiontime1} techniques are widely adopted to improve the estimation accuracy of channel state information (CSI). Meanwhile, using machine learning methods to tackle non-traditional problems in communications has become a new technology trend \cite{DLTELECOM1}\cite{DLTELECOM2}. Recently, as an innovative and efficient branch of the model-free machine learning methods, extreme learning machine (ELM) has gathered much interest from researchers in diversified areas. Owing to the unique properties such as fast training, solution uniqueness, and good generalization ability, ELM is promising to handle the channel interpolation tasks. However, the standard ELM was originally proposed to process vectorized data \cite{elmbasic}, which is not directly applicable to address the channel interpolation problem of multiple-input multiple-output (MIMO) channels. Specifically, MIMO channels exhibit frequency-space correlations, which are often recorded in the form of matrix or tensor, for which direct vectorization will lead to information loss. There have been attempts that tackles the channel interpolation problem with a tensor structure as in \cite{tensorchaninterpol1}, a tensor filter is adopted for a 2D interpolation task. In \cite{tensorchaninterpol2}, the MMSE method has been extended for similar task using tensor technique. To better resolve the high dimensional channel interpolation problem, a novel tensor based ELM, which is capable of handling tensorial inputs and learning through the CSI in MIMO channels, is needed.

In general, there have been great efforts in adapting ELM to tensorial inputs by applying certain matrix/tensor decomposition techniques\cite{tuckerelm!}, which are usually empirical. In this paper, we propose a novel ELM model with tensorial inputs (TELM) to extend the traditional ELM models for tensorial contexts while retaining the valuable ELM features.
Moreover, we further propose a Tucker decomposed extreme learning machine (TDELM) based on the Tucker decomposition method \cite{TuckerDecomposition} to reduce the computational complexity and establish a theoretical argument for its interpolation capability accordingly. Experimental results verify that our proposed methods can achieve comparable performance against the traditional ELMs but with reduced complexity, and outperform the other methods.

The remainder of this paper is organized as follows. Section \ref{sec:preliminaries} reviews the background of single-layer feedforward neural networks (SLFNs), traditional ELM, tensor operations and Tucker decomposition. Section \ref{sec:framework} presents the proposed TELM and TDELM models and discusses how they will be applied to channel interpolation, Section \ref{sec:interpolation} investigates the properties of the considered models, and Section \ref{sec:simulation} demonstrates the experimental results. Finally, Section \ref{sec:conclusion} concludes the paper.
\section{Preliminaries}
\label{sec:preliminaries}
\subsection{Single-layer Feedforward Networks with Vector Inputs}
Consider a dataset with $N$ data samples ($\mathbf{x}_{i},t_{i}$) for $i=1,2,\ldots,N$, where $\mathbf{x}_i\in \mathbb{R}^M$ is the feature vector of the \mbox{$i$-th} sample and $t_i$ is its label. Assume that an SLFN \cite{elmbasic} contains $M$ input neurons and $L$ hidden neurons. The prediction $o_i$ of the label $t_i$ can be formulated as
$o_{i}=\sum_{j=1}^{L}\beta_{j}\sigma(\mathbf{w}_{j}^T\mathbf{x}_{i}+b_{j}),$
where $\mathbf{W}=(\mathbf{w}_1,\mathbf{w}_2,\ldots,\mathbf{w}_L)$ is the weight matrix, whose \mbox{$(i,j)$-th} entry $w_{i,j}$ is the weight between the \mbox{$i$-th} input neuron and the \mbox{$j$-th} hidden neuron; $\mathbf{b}=(b_1,b_2,\ldots,b_L)^T$ is the bias vector from the input layer to the \mbox{$j$-th} hidden neuron; $\boldsymbol{\beta}=(\beta_1,\beta_2,\ldots,\beta_L)^T$ is the weight vector between the hidden layer and the output neuron; and $\sigma$ is the sigmoid activation function, defined as $\sigma(x)\triangleq 1/(1+e^{-x})$. In this setting, the bias between the hidden layer and the output layer is omitted. We then aim to solve the following optimization problem
\begin{equation}
\min_{\mathbf{W},\mathbf{b},\boldsymbol{\beta}} \quad f(\mathbf{W},\mathbf{b},\boldsymbol{\beta})=||\mathbf{T}-\mathbf{O}||_2^2=\sum_{i=1}^N (t_i-o_i)^2,
\end{equation}
where $\mathbf{T}=(t_1,t_2,\ldots,t_N)^T$ and $\mathbf{O}=(o_1,o_2,\ldots,o_N)^T$ are the label vector and its prediction vector, respectively. A typical algorithm finds the optimal values of $\mathbf{W},\mathbf{b}$ and $\boldsymbol{\beta}$ by propagating the errors backwards utilizing gradient or sub-gradient descent methods. However, the algorithm can be very sensitive to initialization and might be stuck at a local minimum due to the fact that the objective function is often non-convex. In addition, the algorithm can be time-costing, which restricts its usage in practical applications.

\subsection{Traditional Extreme Learning Machines with Vector Inputs}
Traditional ELMs are originally designed to train SLFNs\cite{elmbasic}, which improve the training speed remarkably by randomly assigning $\mathbf{w}$ and $\mathbf{b}$, transferring the aforementioned optimization into least square problems. More specifically, let $\mathbf{H}$ be an $N\times L$ matrix, whose \mbox{$(i,j)$-th} entry is $\sigma(\mathbf{w}_j^T \mathbf{x}_i + b_j)$. Instead of minimizing the objective function with respect to $\mathbf{W}$, $\mathbf{b}$, and $\boldsymbol{\beta}$, each entry of $\mathbf{W}$ and $\mathbf{b}$ is drawn from a continuous random distribution. After the matrix $\mathbf{H}$ is calculated, the solution to $\min_{\boldsymbol{\beta}} ||\mathbf{H}\boldsymbol{\beta}-\mathbf{T}||_2^2$ is given by $\hat{\boldsymbol{\beta}}=\mathbf{H}^\dagger \mathbf{T}$ according to the Gauss-Markov theorem \cite{meyerbook},
where $\mathbf{H}^\dagger$ is the Moore-Penrose pseudoinverse of $\mathbf{H}$.
\subsection{Tensor Operations and Tucker Decomposition}
In this paper, we treat tensors as multi-dimensional arrays. Specifically, a tensor $\mathbf{X} \in \mathbb{R}^{I_1 \times I_2 \times \cdots \times I_K}$ of order $K$ stores $I_1\times I_2 \times \cdots \times I_K$ elements. 
The inner product of two tensors, e.g., $\mathbf{X}, \mathbf{Y}\in \mathbb{R}^{I_1 \times I_2 \times \cdots \times I_K}$, is defined as $\langle \mathbf{X},\mathbf{Y} \rangle = \langle \textrm{vec}(\mathbf{X}),\textrm{vec}(\mathbf{Y}) \rangle$. The vectorization $\textrm{vec}(\mathbf{X})$ of a tensor $\mathbf{X}$ is obtained by stacking the elements of $\mathbf{X}$ into an $\left(\prod_{k=1}^K I_k\right)$-dimensional column vector in a fixed order \cite{tensorreviewk},  e.g., in a lexicographical order or reverse lexicographical order.
Tensors can also be unfolded into matrices. Given a tensor $\mathbf{X} \in \mathbb{R}^{I_1 \times I_2 \times \cdots \times I_K}$ and an index $k$, the $k$-mode matricization $\mathbf{X}_{(k)}$ of $\mathbf{X}$  is a matrix with $\prod_{j\ne k} I_j$ columns and $I_k$ rows obtained by unfolding $\mathbf{X}$ along the \mbox{$k$-th} coordinate\cite{tensorreviewk}. For $1\le k\le K$, the $k$-mode rank of tensor $\mathbf{X}$, denoted by $\textrm{rank}_{k}(\mathbf{X})$, is defined as the rank of $\mathbf{X}_{(k)}$, which satisfies $\textrm{rank}_k(\mathbf{X})\le I_k$\cite{tensorreviewk}.

The Tucker decomposition is a branch of the higher-order singular value decomposition \cite{TuckerDecomposition}. Consider $\mathbf{X} \in \mathbb{R}^{I_1 \times I_2 \times \cdots \times I_K}$ and a vector $(D_1,D_2,\ldots,D_K)$. If $D_k\ge \textrm{rank}_k(\mathbf{X}$) for all $k\in \{1,2,\ldots,k\}$, there exist a core tensor $\mathbf{X}'\in \mathbb{R}^{D_1 \times D_2 \times \cdots \times D_K}$ and $K$ factor matrices, \emph{i.e.,} $\mathbf{B}(1),$ $\mathbf{B}(2),$ $\ldots,$ $\mathbf{B}(K)$ with each $\mathbf{B}(k)\in \mathbb{R}^{I_k \times D_k}$, to be column-wise orthogonal: 
\begin{align*}\label{tuckereq}
	\mathbf{X}_{i_1 i_2 \cdots i_K}=&\ \sum_{d_1=1}^{D_1}\sum_{d_2=1}^{D_2}\cdots \sum_{d_K=1}^{D_K}\mathbf{X}'_{d_1 d_2 \cdots d_K}\Bigg[\prod_{j=1}^K\mathbf{B}(j)_{i_j,d_j}\Bigg],
\end{align*}
which could be written compactly as $\mathbf{X} =\llbracket\mathbf{X}';\mathbf{B}(1),\ldots,\mathbf{B}(K)\rrbracket$. If $D_K=I_K$, there exists a core tensor $\mathbf{X}'\in \mathbb{R}^{D_1 \times D_2 \times \cdots \times D_K}$ and $K$ factor matrices $\{\mathbf{B}(k)\}_{k=1}^K$ with $\mathbf{B}(k)\in \mathbb{R}^{I_k \times D_k}$ for $1\le k\le K-1$ and $\mathbf{B}(K)=\mathbf{I}$ (an $I_K\times I_K$ identity matrix) such that $\mathbf{X} =\llbracket\mathbf{X}';\mathbf{B}(1),\ldots,\mathbf{B}(K-1),\mathbf{I}\rrbracket$. If $D_k<\textrm{rank}_k(\mathbf{X})$ for some $k$, such a core tensor and factor matrices do not exist. As an alternative, we can use an approximation of $\mathbf{X}$ by the truncated Tucker decomposition \cite{tensorreviewk}.

We next introduce an important property of the Tucker decomposition mentioned in \cite{tts}.
\begin{lem}[Duality Lemma]\label{Duality Lemma}
	Given a tensor pair $\mathbf{W}\in  \mathbb{R}^{I_1\times I_2\times \cdots \times I_K}$ and $\mathbf{X}\in  \mathbb{R}^{I_1\times I_2\times \cdots \times I_K}$, and their corresponding tucker decomposed core tensor $\mathbf{W'}$ and $\mathbf{X'}$, where $\mathbf{W} =\llbracket\mathbf{W}';\mathbf{B}(1),\ldots,\mathbf{B}(K)\rrbracket$ and $\mathbf{X} =\llbracket\mathbf{X}';\mathbf{B}(1),\ldots,\mathbf{B}(K)\rrbracket$, we have $\langle \mathbf{W}',\mathbf{X}'\rangle=\langle \mathbf{W},\mathbf{X}\rangle$.
\end{lem}

Lemma \ref{Duality Lemma} tells us that the inner product can be done on the tucker decomposed core pair instead of on the original pair without loss in accuracy. If the original tensor admits a low-rank structure, the computational cost could be drastically reduced by calculating the inner product of the core tensors. 

\begin{table}[t]
	\begin{center}
		\begin{tabular}{ lc }
			\hline\hline
			\textbf{Algorithm 1 TELM} \\
			\hline
			\noindent \textbf{Input:} Data samples $(\mathbf{X}_i,t_i)$, $i=1,2,\ldots,N$.\\
			\noindent \textbf{Output:} Weight vector $\boldsymbol{\beta}$.\\
			1 \ Draw each entry of the weight tensor $\mathcal{W}$ and the bias\\
			\ \ \ \hspace{-2pt} vector $\mathbf{b}$ from a continuous random distribution.\\
			2 \ Calculate matrix $\mathbf{H}$ from $\{\mathbf{X}_i\}_{i=1}^N$, $\mathcal{W}$, and $\mathbf{b}$ by (\ref{defH}).\\
			3 \ Calculate parameter vector $\boldsymbol{\beta}$ by $\hat{\boldsymbol{\beta}}=\mathbf{H}^\dagger \mathbf{T}.$\\
			4 \ Return $\boldsymbol{\beta}$.\\
			\hline\hline
		\end{tabular}       \vspace{-0.5cm}
	\end{center}
\end{table}

\section{Frameworks}\label{sec:framework}

\subsection{Channel Interpolation}
To conduct high-efficiency data transmission in a typical Multi-input Multi-output Orthogonal Frequency Division Multiplexing (MIMO-OFDM) setting, CSI for each sub-carrier is required. One way of acquiring CSI is inserting a pilot signal to each sub-carrier, and then calculating the CSI at receiver. To further reduce the overhead of probing, the following interpolation scheme is adopted. Let $S_i$ denote the $i$-th sub-carrier's CSI; pilots are inserted into sub-carriers with odd indices. The CSI of sub-carriers with even indices are then inferred as
\begin{equation*}
	S_i=f(S_{i-w+1},\ldots,S_{i-3},S_{i-1},S_{i+1},S_{i+3},\ldots,S_{i+w-1}),
\end{equation*}
where $w$ is the window length and $f$ is the interpolation function. In our work, we adopt the modified ELM as $f$ and the detail will be explained in the next section. Notice that different window design can be adopted to achieve the balance between carrier usage and accuracy.
\label{sec:proposed_model}
\subsection{Extreme Learning Machines with Tensorial Inputs}
Consider a dataset with $N$ data samples $(\mathbf{X}_i,t_i)$ for $i=1,2,\ldots,N$, where $\mathbf{X}_i\in \mathbb{R}^{I_1\times I_2\times \cdots \times I_K}$ is a tensor of order $K$, and $t_i$ is its label. For an SLFN with $I_1\times I_2\times \cdots \times I_K$ input neurons and $L$ hidden neurons, its prediction $o_i$ of label $t_i$ could be calculated as
$o_{i}=\sum_{j=1}^{L} \beta_{j} \sigma(\langle \textrm{vec}(\mathbf{W}_{j}), \textrm{vec}({\mathbf{X}}_{i}) \rangle+b_{j})$, where $\mathcal{W}=(\mathbf{W}_1,\mathbf{W}_2,\ldots,\mathbf{W}_L)$ with $\mathbf{W}_j$ as the weight tensor of the \mbox{$j$-th} hidden neuron; $\mathbf{b}$, $\boldsymbol{\beta}$, $\mathbf{T}$, $\mathbf{O}$, and $\sigma$ are defined as in Section II. The goal is still to minimize
$f(\mathcal{W},\mathbf{b},\boldsymbol{\beta})=||\mathbf{T}-\mathbf{O}||_2^2=\sum_{i=1}^N (t_i-o_i)^2.$
Consistent with the traditional ELM, we draw each entry of the weight tensor $\mathcal{W}$ and the bias vector $\mathbf{b}$ from a continuous random distribution and solve the problem $\min_{\boldsymbol{\beta}} ||\mathbf{H}\boldsymbol{\beta}-\mathbf{T}||_2^2$, where
\begin{equation}\label{defH}
\mathbf{H}=\left[
\begin{array}{ccc}
\sigma(\langle \mathbf{W}_1,\mathbf{X}_1\rangle + b_1)&\dots&\sigma(\langle \mathbf{W}_{L},\mathbf{X}_1\rangle + b_{L})\\
\vdots&\ddots&\vdots\\
\sigma(\langle \mathbf{W}_1,\mathbf{X}_N\rangle + b_1)&\dots&\sigma(\langle \mathbf{W}_{L},\mathbf{X}_N\rangle + b_{L})\\
\end{array}
\right].
\end{equation}
The problem has a unique least square solution $\hat{\boldsymbol{\beta}}=\mathbf{H}^\dagger \mathbf{T}$, with $\mathbf{H}^\dagger$ is the Moore-Penrose pseudoinverse of $\mathbf{H}$ as defined above. Detailed procedures are summarized in Algorithm 1.
It is noteworthy to point out that the TELM handles the tensorial inputs with the same computational cost as the traditional ELM with vectorized inputs.
\subsection{\mbox{Tucker Decomposed Extreme Learning Machines}}
To improve the learning efficiency of TELM, we further propose the Tucker decomposed extreme learning machine (TDELM) based on the Tucker decomposition method. Generally, computing $\mathbf{H}$ in (\ref{defH}) requires $NL\prod_{i=k}^K I_k$ multiplication operations, which is computationally demanding when dealing with a large dataset. However, by employing the Tucker decomposition, the computational cost of computing $\mathbf{H}$ could be largely reduced when working with the datasets with a low $ k $-mode rank.

Consider a dataset with $N$ samples $(\mathbf{X}_i,t_i)$ for $i=1,2,\ldots,N$. We first concatenate $\mathbf{X}_1,\mathbf{X}_2,\ldots, \mathbf{X}_N$ into a tensor $\mathbf{X}\in \mathbb{R}^{I_1\times I_2\times \cdots \times I_K\times N}$ of order $(K+1)$, then find the $k$-mode rank $n_k$ of $\mathbf{X}$ for $1\le k\le K$, and next apply Tucker decomposition to $\mathbf{X}$ such that the core tensor $\mathbf{X'}$ is of size $n_1\times n_2\times \ldots\times n_K\times N$ and the \mbox{$(K+1)$-th} factor matrix $\mathbf{B}(K+1)$ is an identity matrix. Afterwards, we extract $\mathbf{X'}$ along the \mbox{$(K+1)$-th} axis into $N$ subtensors $\mathbf{X'}_1,\mathbf{X}'_2,\ldots,\mathbf{X}'_N$. We next consider an SLFN with $n_1\times n_2\times \cdots \times n_K$ input neurons and $L$ hidden neurons. Let $\mathcal{W}'=(\mathbf{W}'_1,\mathbf{W}'_2,\ldots,\mathbf{W}'_L)$ be the total weight tensor with $\mathbf{W}'_j$ as the weight tensor of the \mbox{$j$-th} hidden neuron, $\mathbf{b}'=(b'_1,b'_2,\ldots,b'_L)^T$ be the bias scalar from the input layer to the \mbox{$j$-th} hidden neuron, and $\boldsymbol{\beta}'=(\beta'_1,\beta'_2,\ldots,\beta'_L)^T$ be the weight vector between the hidden layer and the output neuron. Each entry of $\mathcal{W}'$ or $\mathbf{b}'$ is randomly drawn from a continuous random distribution, and $\mathbf{H}$ is then calculated as
\begin{equation}\label{deftuckerH}
\mathbf{H}=\left[
\begin{array}{ccc}
\sigma(\langle \mathbf{W}'_1,\mathbf{X}'_1\rangle + b_1)&\dots&\sigma(\langle \mathbf{W}'_{L},\mathbf{X}'_1\rangle + b_{L})\\
\vdots&\ddots&\vdots\\
\sigma(\langle \mathbf{W}'_1,\mathbf{X}'_N\rangle + b_1)&\dots&\sigma(\langle \mathbf{W}'_{L},\mathbf{X}'_N\rangle + b_{L})\\
\end{array}
\right].
\end{equation}
Finally, we solve the optimization problem
$\min_{\boldsymbol{\beta}} ||\mathbf{H}\boldsymbol{\beta}-\mathbf{T}||_2^2,$
with the least square square solution $\hat{\boldsymbol{\beta}}=\mathbf{H}^\dagger \mathbf{T}$. The above procedures are summarized in Algorithm 2.
\begin{table}[!h]\label{TDELMalgorithm}
	\begin{center}
		\begin{tabular}{ lc }
			\hline\hline
			\textbf{Algorithm 2 TDELM} \\
			\hline
			\noindent \textbf{Input:} Data samples $(\mathbf{X}_i,t_i)$, $i=1,2,\ldots,N$.\\
			\noindent \textbf{Output:} Weight vector $\boldsymbol{\beta}$.\\
			1 \ Concatenate $\mathbf{X}_1,\hspace{-0.2mm}\ldots,\hspace{-0.3mm} \mathbf{X}_N$ into a tensor $\mathbf{X}$ of order $(K\hspace{-0.80mm}+\hspace{-0.70mm}1)$.\!\!\!\!\!\\
			2 \ Find the $k$-mode rank $n_k$ of $\mathbf{X}$ for $1\le k\le K$.\\
			3 \ Conduct a Tucker decomposition such that \\
			\ \ \ \hspace{-2pt} $\mathbf{X} =\llbracket\mathbf{X}';\mathbf{B}(1),\ldots,\mathbf{B}(K),\mathbf{I}\rrbracket$, $\mathbf{X}'\in \mathbb{R}^{n_1 \times \cdots \times n_K\times N}$,\\
			\ \ \ \hspace{-2pt} $\mathbf{B}(k)\in \mathbb{R}^{I_k \times n_k}$ for $1\le k\le K$, and $\mathbf{I}\in \mathbb{R}^{N \times N}$.\\
			4 \ Draw each entry of weight matrix $\mathcal{W}'$ and bias vector $\mathbf{b}'$\!\!\!\!\!\\
			\ \ \ \hspace{-2pt} from a continuous random distribution.\\
			5 \ Calculate matrix $\mathbf{H}$ from $\mathbf{X}'$, $\mathcal{W}'$, and $\mathbf{b}'$ by (\ref{deftuckerH}).\\
			6 \ Calculate parameter vector $\boldsymbol{\beta}$ by $\hat{\boldsymbol{\beta}}=\mathbf{H}^\dagger \mathbf{T}.$\\
			7 \ Return $\boldsymbol{\beta}$.\\
			\hline\hline
		\end{tabular}\vspace{-0.5cm}
	\end{center}
\end{table}


\section{Properties of TELM/TDELM}
\label{sec:interpolation}
In this section, we establish the interpolation theorems for the TELM and TDELM.
Specifically, given a dataset with $N$ distinct data samples $(\mathbf{X}_i,t_i), \ i=1,2,\ldots,N$, a TELM or TDELM with $N$ hidden neurons and sigmoid activation functions has the properties as follows.

\begin{thm}[Interpolation Capability of TELMs]\label{theoreminterpolation}
	Assume that each entry of the weight tensor $\mathcal{W}$ and bias vector $\mathbf{b}$ is randomly chosen from an interval according to a continuous probability distribution. Then with probability one, $\mathbf{H}$ in (\ref{defH}) is invertible and $||\mathbf{H}\boldsymbol{\beta}-\mathbf{T}||_2=0$.
\end{thm}

\begin{thm}[Interpolation Capability of TDELMs]\label{theoreminterpolationtucker}
	Assume that each entry of the weight tensor $\mathcal{W}'$ and bias vector $\mathbf{b}'$ is randomly chosen from an interval according to a continuous probability distribution. Then with probability one, $\mathbf{H}$ in (\ref{deftuckerH}) is invertible and $||\mathbf{H}\boldsymbol{\beta}-\mathbf{T}||_2=0$.
\end{thm}
Theorem \ref{theoreminterpolation} can be treated as a special case of Theorem \ref{theoreminterpolationtucker} by setting $\mathbf{X'}=\mathbf{X}$ and $\mathbf{B}(k)=\mathbf{I}$ for $1\le k\le K$, the proof of Theorem \ref{theoreminterpolationtucker} is provided as follows.



\textit{Proof:} Define $\boldsymbol{\rho}_j(y)=[\rho_{j,1}(y),\ldots,\rho_{j,N}(y)]^T$, where $\rho_{j,i}(y)=\sigma(\langle \mathbf{W}'_j,\mathbf{X}'_i\rangle + y)$ for $1\le i\le N$. Note that $\boldsymbol{\rho}_j(b_j)$ is the \mbox{$j$-th} column of $\mathbf{H}$ in (\ref{deftuckerH}). Let $\psi_k=\langle\mathbf{W}'_j,\mathbf{X}'_k\rangle$ for $1\le k\le N$. Each entry of $\mathbf{W_j}'$ is drawn from a continuous distribution over an interval and $\mathbf{X'}_i$s are distinct from each other; thus $\langle \mathbf{W}'_j,\mathbf{X}'_k\rangle\ne \langle \mathbf{W}'_j,\mathbf{X}'_l\rangle$ for $k\ne l$ with probability one. Therefore, $\psi_1,\psi_2,\ldots,\psi_N$ are distinct from each other with probability one.

Now assume that $\boldsymbol{\rho}_j(y)$ belongs to a subspace of dimension less than $N$. Then there will be a vector $\boldsymbol{\gamma}=(\gamma_1,\gamma_2,\ldots,\gamma_N)^T$ that is orthogonal to this subspace. Therefore,
$\langle \boldsymbol{\gamma}, \boldsymbol{\rho}_j(y) \rangle=0$ for all $y$, \emph{i.e.,} $\sum_{i=1}^N \gamma_i \sigma(y+\psi_i)=0.$
We will then have
$S(j)\triangleq\sum_{i=1}^N\gamma_i \sigma^j(y+\psi_i)=0$
for $j=1,2,\ldots,N$ due to the fact that for $1\le j\le N-1$, taking the derivative of $S(j)=0$ on both sides over $y$ yields
$j\sum_{i=1}^N\gamma_i \sigma^j(y+\psi_i) [1-\sigma(y+\psi_i)]=0,$
and hence
$S(j+1)=S(j)-\sum_{i=1}^N\gamma_i \sigma^j(y+\psi_i) [1-\sigma(y+\psi_i)]=0.$
Let $\mathbf{G}$ be an $N\times N$ matrix whose \mbox{$(i,j)$-th} entry is $\sigma^i(y+\psi_j)$ for $1\le i,j\le N$.
Then
$\mathbf{G}\cdot (\gamma_1,\gamma_2,\ldots,\gamma_N)^T=(S(1),S(2),\ldots,S(N))^T=\mathbf{0},$
and hence $\mathbf{G}$ is a singular matrix.

On the other hand,
\begin{align*}
&\det(\mathbf{G})=\sigma(y+\psi_1)\sigma(y+\psi_2)\cdots \sigma(y+\psi_N)\cdot \det(\mathbf{G}')\\
&=\prod_{k=1}^N\sigma(y+\psi_k)\cdot \prod_{1\le i<j\le N} [\sigma(y+\psi_j)-\sigma(y+\psi_i)]\ne 0,
\end{align*}
where $\mathbf{G}'$ is an $N\times N$ matrix whose \mbox{$(i,j)$-th} entry is $\sigma^{i-1}(y+\psi_j)$ for $1\le i,j\le N$, the second equality follows from the Vandermonde determinant \cite{vander}, and the last inequality follows from the fact that $\sigma(y+\psi_i)$'s are distinct from each other. Then a contradiction appears. Thus $\boldsymbol{\rho}_j(y)$ belongs to a subspace of dimension $N$, and matrix $\mathbf{H}$ is thus of full rank with probability one. \hfill $\blacksquare$

\section{Simulation Results}
\label{sec:simulation}
In this section, an experiment is conducted over a real-world wireless MIMO channel response dataset to compare the performance of training time and accuracy among multiple methods including the purposed TDELM. We will show that: 1) TDELM achieves comparable prediction accuracy against other methods and 2) TDELM requires lower computational cost and shorter training time than ELM and SLFN over decomposed data.

The dataset was generated via conducting experiments in a lecture hall. Specifically, a 64-element virtual uniform linear array (ULA) is used as the transmitter (Tx) antenna array, and the receiver (Rx) with a single antenna is deployed at three different locations within the auditorium of the hall. The carrier frequency is set to be 4 GHz, and the measurement bandwidth is 200 MHz that is uniformly sampled at 513 frequency points. Sixteen continuous snapshots are obtained for each sub-channel. The obtained results are stored in a tensor of size $64\times 3\times 8208$, and each entry represents the response from one Tx array element to one Rx array element at a given frequency point. More detailed descriptions about the dataset could be found in \cite{datasource}.

The channel responses are normalized with zero-mean and unit standard derivation. The window size $W$ is set to be 4. The feature tensor $\mathbf{X}\in \mathbb{R}^{64\times3\times 4 \times 8208}$ is created by a sliding window method, and the $(j,k,w,l)$-th element of $\mathbf{X}$ is given by $X_{j,k,w,l}=H_{jk(w-W+1+2(l-1))}$. The sliding window method will capture the local correlation in the neighborhood of adjacent points and the tensor decomposition is conducted on tensor $\mathbf{X}$.

The dataset is divided evenly into two subsets, one as the training set and the other as the test set, both containing 4,104 samples. Two neural networks, an ELM and a TDELM, are constructed with the same hidden-layer neurons for comparison fairness. The Tucker decomposition was implemented by modifying the Tensor Toolbox library \cite{ttoolbox}. Grid search is conducted for the hyperparameters: the net size and the decomposition size. The results of mean squared error (MSE) and running time are recorded. To mitigate the fluctuation caused by the randomness, the SelectBest method \cite{selectbest} is adopted and training is repeated 100 times to find the best parameter set. Least mean square filtering (LMSE), where the weighted combination of the data over a window is used as the prediction and each sub channel has its own weight parameter, and input averaging (Mean), where the mean of data over a window is used as the prediction, are also used as the basic comparison schemes.

\begin{figure}
	\centering
	\includegraphics[width=1\linewidth]{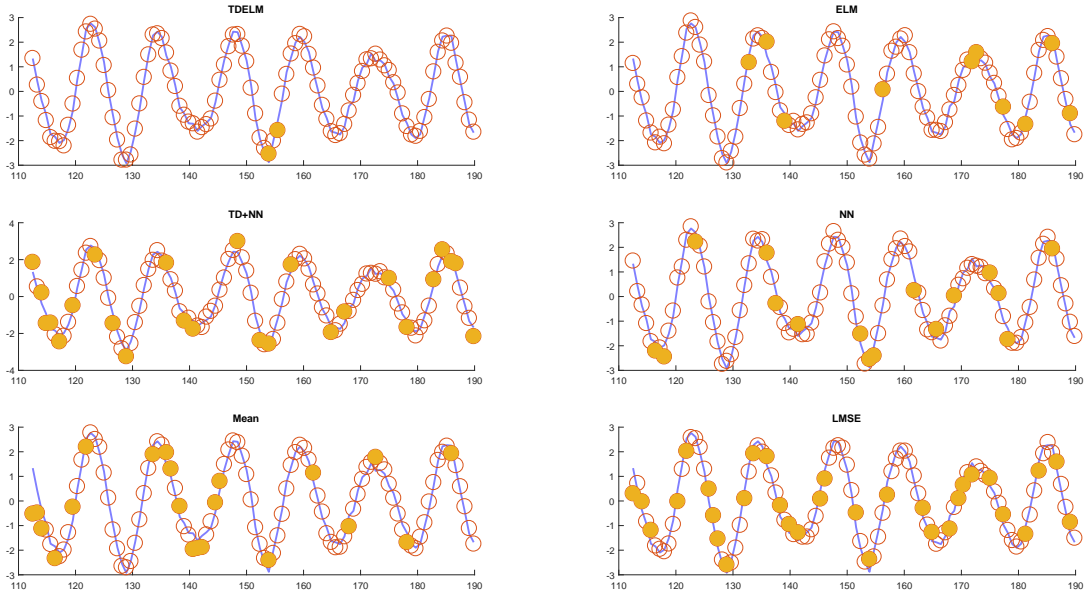}
	\caption{Prediction results of learning machines}
	\setlength\abovecaptionskip{0pt}
	\setlength\belowcaptionskip{0pt}
	\label{fig:origrecovpart}
\end{figure}
\begin{table}[]
	\centering
	\setlength\abovecaptionskip{0pt}
	\setlength\belowcaptionskip{0pt}
	\caption{MSE and time consumption (TC)}
	\label{table:1}
	\begin{tabular}{|c|c|c|c|}
		\hline
		&TDELM  &ELM  &TD+NN    \\ \hline
		MSE	&0.0280&0.0286&0.0645\\ \hline
		TC (s)&1.3786  &1.5896  & 397.28 \\ \hline
		&NN  &Mean  &LMSE\\ \hline
		MSE	&0.0363&0.0769&0.0377\\ \hline
		TC (s)	&689.94  & 0.0334  & 0.0805  \\ \hline
	\end{tabular}
\end{table}
The choices of hyperparameters are as follows. The node size in the hidden layer is 1,080, and the decomposition shrinks each tensor from $\{64,3,4\}$ to $\{64,2,2\}$, which leads to 66\% less multiplications needed per inner production operation. The Tucker decomposition acquired by the Alternative Least Square algorithm\cite{ttoolbox} costs about 6 seconds for this dataset. The decomposition time does not scale with repeated training thus it is excluded from Table 1. The performance in terms of MSE and time consumption (TC) are summarized in Table 1 and a snapshot of the interpolation results has been shown in Figs. 1 and 2.

\begin{figure}
	\centering
	\includegraphics[width=1\linewidth]{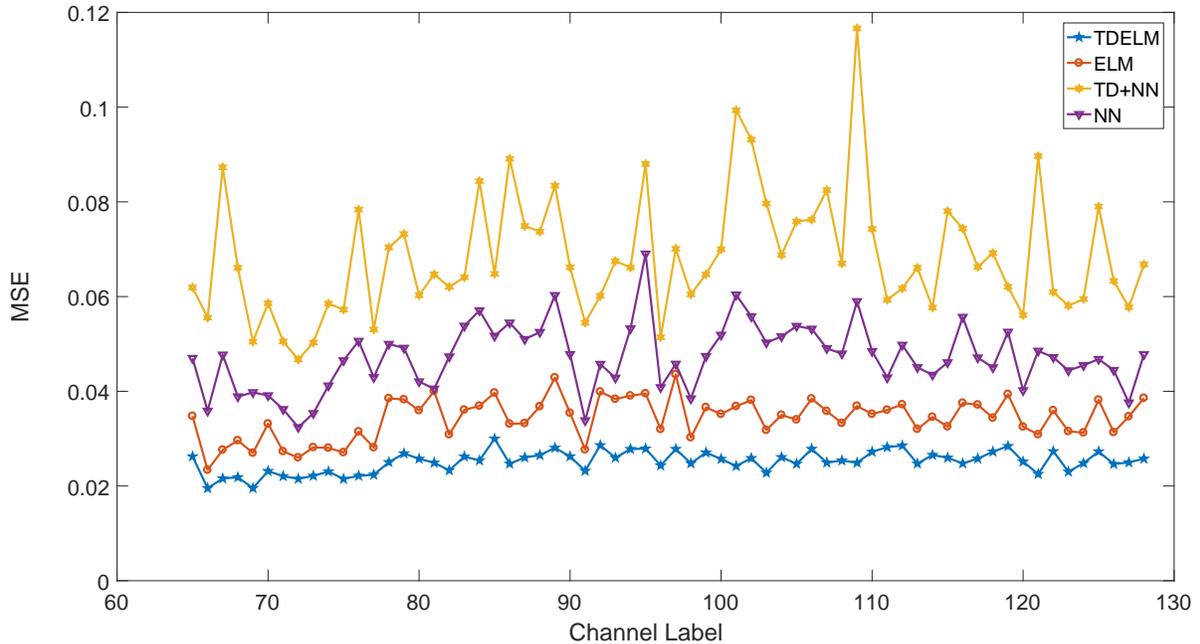}
	\caption{Prediction results of learning machines.}
	\setlength\abovecaptionskip{0pt}
	\setlength\belowcaptionskip{0pt}
	\label{fig:mse}
\end{figure}

From Table \ref{table:1} and Fig.~\ref*{fig:mse}, it is shown that overall, TDELM achieved consistent gains in terms of MSE compared to all other methods, while also being the fastest on the decomposed dataset, compared with the TD+NN method. To showcase the gain better, one instance of the curve is shown in Fig. \ref{fig:origrecovpart}, where the blue lines denote the true values and yellow markers denote the interpolated values, respectively. When the prediction error for a particular point exceeds  0.3, which is roughly 10\% of the maximal amplitude, the corresponding marker is filled as solid, which indicates a significant error. We can see there are much more significant errors among methods other than TDELM, which corresponds to smaller MSE shown in Fig.~\ref*{fig:mse}. Another interesting observation is, although TDELM has better accuracy than ELM, the TD+NN combination performs the worst. This implies that the gain might not only comes from applying tensor decomposition on the input data, but also from using the corresponding learning machine designed with such decomposition in mind.

\section{Conclusion}
\label{sec:conclusion}

In this paper, we proposed an extreme learning machine with tensorial inputs (TELM)
and a Tucker decomposed extreme learning machine (TDELM) to handle the channel interpolation task in MIMO system. Moreover, we established a theoretical argument for the interpolation capability of TDELM. The experimental results verified that our proposed TDELM can achieve comparable performance against the traditional ELM but with reduced complexity, and outperform the other methods considerably in terms of the channel interpolation accuracy.

\ifCLASSOPTIONcaptionsoff
  \newpage
\fi

\end{document}